# Temperature Dependence Of Spin Resonance In Cobalt Substituted NiZnCu Ferrites

A. Lucas [1,2], R. Lebourgeois [1], F. Mazaleyrat [2],E. Labouré[2]
1. THALES R&T, Campus Polytechnique, 1 avenue Augustin Fresnel, 91767 Palaiseau, France
2. SATIE, ENS de Cachan, 61 av. du Président Wilson, 94235 Cachan, France

**Abstract** : Cobalt substitutions were investigated in $Ni_{0.4}Zn_{0.4}Cu_{0.2}Fe_2O_4$ ferrites, initial complex permeability were then measured from 1 MHz to 1 GHz. It appears that cobalt substitution led to a decrease of the permeability and an increase of the $\mu_s \times f_r$ factor. As well, it gave to the permeability spectrum a sharp resonance character. We also observed a spin reorientation occuring at a temperature depending on the cobalt content. Study of the complex permeability versus temperature highlighted that the most resonant character was obtained at this temperature. This shows that cobalt contribution to second order magneto-crystalline anisotropy plays a leading role at this temperature.

Nickel-zinc-copper ferrites are interesting materials because of their high permeability in MHz range. Moreover, their low sintering temperatures make them suitable for the realization of integrated components in power electronic. As for nickel-zinc ferrites, cobalt substitution is an efficient technique to decrease permeability [1] and magnetic losses of nickel-zinc-copper ferrites [2]. It has been proposed that the effect of cobalt is to produce pinning of the domain walls because of anisotropy enhancement due $Co^{2+}$ ions ordering [3]. The aim of this paper is to study the effect of cobalt substitution and particularly the role of the cobalt contribution to anisotropy. Ferrites of formula $(Ni_{0.4}Zn_{0.4}Cu_{0.2})_{1-x}Co_xFe_{1.98}O_4$ were studied for cobalt substitutions up to 0.035 mol.

Ferrites were synthesized using the conventional ceramic route. The raw materials ($Fe_2O_3$, NiO, ZnO, CuO) were ball milled for 24h hours in water. $Co_3O_4$ was then added before the calcination at 760°C in air for 2 hours. The calcined ferrite powder was then milled by attrition for 30 min. The resulting powder was compacted using axial pressing. The sintering was performed at 935°C for 2 hours in air. Magnetic characterizations were done on ring shaped samples with the following dimensions: external diameter = 6.8 mm; internal diameter = 3.15 mm; height = 4 mm. Initial complex permeability ($\mu'$ and $\mu''$) was measured versus frequency between 1 MHz and 1 GHz using an impedance-meter HP 4291. Static initial permeability ($\mu_s$) was defined as $\mu'$ at 1 MHz because for these ferrites $\mu'$(1 MHz) = $\mu'$(100 Hz). For permeability versus temperature measurements, the rings were wound with a copper wire and placed in an oven going from –70°C to 200°C. $\mu_s$ was deduced from the inductance measured at 100 kHz by an impedance-meter Agilent 4194A.

The samples sintered at 935°C have all the pure spinel crystalline structure and a density higher than 96% of the theoretical density. Table I shows evolution of the permeability of the $(Ni_{0.4}Zn_{0.4}Cu_{0.2})_{1-x}Co_xFe_{1.98}O_4$ ferrites. Cobalt substitutions lead to a decrease of the initial complex permeability and an increase of the $\mu_s \times f_r$ factor which is maximum for Co = 0.021 mol ($f_r$ is the frequency resonance defined as the maximum of $\mu''$). The raise of this factor shows that $f_r$ increases faster than $\mu_s$ decreases.

Figure 1 shows the initial complex permeability versus frequency for $(Ni_{0.4}Zn_{0.4}Cu_{0.2})_{1-x}Co_xFe_{1.98}O_4$ ferrites with different cobalt content. One can see that the spectra become sharper when the cobalt rate increases. It is accepted that the permeability has two contributions : at low frequency wall domain displacements are preponderant and at higher frequency, permeability is mainly due to the spin rotation [4]. In general, the relaxation behavior of domain walls essentially hides the spin resonance, but cobalt is known to inhibit the domain wall displacements [5], which produces two effect: (i) the initial permeability is decreasing with cobalt content; (ii) at higher frequency, the cobalt seems to promote the spin rotation by shifting the frequency resonance (maximum of $\mu''$) toward higher frequencies. Consequently, the magnetic losses due to domain wall displacements are lowered, leading to a stronger dissymmetry in the shape of $\mu''$ peak. The magnetic losses rise at higher frequency but with a steeper slope.

The cobalt has also an effect on the temperature variation of the permeability. In order to understand this phenomenon, the permeability dependence on temperature has been studied (figure 2).

The cobalt free ferrite (curve A) shows a monotonous increase in this temperature range. In contrast, the behavior is different for the cobalt-substituted ferrites, for which a local maximum in the initial permeability appears. This is the consequence of the magneto-crystalline anisotropy compensation due to the cobalt ions contribution. Indeed, the first order anisotropy constant of the Ni(ZnCu) ferrite host crystal is negative, whereas Co ferrite has a positive one. As previously described by Van Den Burgt [6] for a certain amount of Co within the order of 0.1/u.f., it results that a spin reorientation transition occurs (SRT, i.e. a change in easy axis) at a temperature $T_0$, increasing with cobalt content. The permeability is described by the following relation :

$$\mu' \ \alpha \ \frac{M_s^{\ 2}}{K_{eff}} \quad [6]$$

$M_s$ is the saturation magnetization and $K_{eff}$ the effective anisotropy. $K_{eff}$ consists of three components due to : magneto-crystalline anisotropy ($K_1$) of the host crystal, the cobalt ions contribution to anisotropy, higher order contributions and magneto-elastic energy [8]. This spin reorientation leads to an increase of the permeability characterised by a local maximum around $T_0$. Below the SRT, $K_1>0$ and $K_2>0$ as the Co contribution dominates, and above SRT $K_1<0$, corresponding to a change from [100] to [110] easy axis. Figure 2, shows that $(Ni_{0.40}Zn_{0.40}Cu_{0.20})_{0.979}Co_{0.035}Fe_{1.98}O_4$ ferrite has a SRT close to room temperature. The strong resonant character of the permeability at this point could be explained by the strong pinning of domain wall due to high second order anisotropy contribution.

In order to go deeper insight the effect of cobalt on the magnetic behavior, the initial complex permeability spectra were recorded near the SRT on two ferrites :

- a $(Ni_{0.4}Zn_{0.4}Cu_{0.2})_{0.965}Co_{0.035}Fe_{1.98}O_4$ ferrite, which has a SRT around 10°C.

- a $Ni_{0.4}Zn_{0.4}Cu_{0.2}Fe_{1.98}O_4$ ferrite which doesn't exhibits SRT.

Figure 3 shows μ'(f) spectra between −50°C and 180°C of these two materials . To quantify the resonant character of the μ' spectrum versus frequency, we defined the following resonance factor $F_{res}$ :

$$F_{res} = \frac{(\mu'_{max} - \mu'_{static})}{\mu'_{static}} \times 100$$

which variation for the two compositions is shown on figure 4.

The cobalt free ferrite has a low $F_{res}$ (around 15%) slightly decreasing with temperature. This behavior can be explained by the decrease of the magnetization saturation. In contrast, the permeability spectra of the cobalt-substituted ferrite strongly depend on the temperature. Figures 3 and 4 show that the sharpest resonant character is obtained around 10°C, corresponding to the SRT. It can also be noted that, in this range of temperature, the resonance factor of the cobalt-substituted ferrites is always higher than one of the cobalt free ferrite. As the magnetization is constantly decreasing in this temperature range, such a change in the permeability behaviour is necessarily due to a change in the effective anisotropy.

In the case of Co substituted ferrite, the resonance should be explained if one considers that domain wall are pinned (case of strong anisotropy) and the spins rotates freely (case of vanishing anisotropy), so there is an apparent contradiction. However, near the SRT, the situation is much different compared to usual: $K_1$ is vanishing but $K_2$ is not necessarily dropping to zero accordingly. In the limit of $K_1 = 0$, there arte two main consequences. Firstly, the domain wall energy is not vanishing but $\gamma = 2(AK_2)^{1/2}$ [9], so the pinning energy may be still important. Secondly, development of anisotropy energy reduces to $\Delta E_A = K_2 \alpha_1^2 \alpha_2^2 \alpha_3^2$. If one of the direction cosines is null (case of {100} and equivalent planes) there is no anisotropy variation. So, a possible explanation of the strong resonance observed only in the cobalt-substituted sample would be nearly free spin rotation in the {100}, {010} and {001} planes [10]. If the magnetization turns in the {110} from $\langle 1 1 \cdot \rangle$ to $\langle 111 \rangle$, the anisotropy change is at most $K_2/8$. As the static permeability is relatively small, this would mean that $K_2$ is relatively strong in this material near the SRT, explaining also why the $\mu_S(T)$ maximum is relatively smooth

In conclusion, the study of permeability spectra of cobalt substituted NiZnCu ferrites as a function of temperature shows that cobalt substitution can shift the spin reorientation transition close to room temperature due to cancellation of first order anisotropy constants of Co ion and the host crystal. As a consequence of the non-zero second order anisotropy constant, spin resonance damping is very small resulting in a strong resonance revealed by a very large overshoot (60%) of μ' spectra and a sharp absorption peak in μ" spectra in the vicinity of SRT.

**References**
[1] T. Y. Byun, S. C. Byeon, K. S. Hong, Factors affecting initial permeability of Co-substituted Ni-Zn-Cu ferrites, IEEE, vol 35, Issue 5, Part 2, pages : 3445-3447, sept 99
[2] R. Lebourgeois, J. Ageron, H. Vincent and J-P. Ganne, low losses NiZnCu ferrites (ICF8), Kyoto and Tokyo, Japan 2000
[3] J. G. M. De Lau, A. Broese van Groenoul, Journal de Physique 38, 1977, page Cl-17
L. Néel, J. Phys. Radium 13, 249, 1952
[4] T. Tsutaoka, M. Ueshima and T. Tokunaga, J. Appl. Phys. 78 (6), p 3983-3991, 1995
[5] A.P. Greifer, V. Nakada, H. Lessoff, J. Appl. Phys., vol 32, 382-383, 1961
[6] C. M. van der Burgt, Philips Research Report 12, 97-122, 1957
[7] H. Pascard, SMM13, J. Phys. IV France, 8, 1998
[8] A. Globus and P. Duplex, J. Appl. Phys., vol 39, no.2 (part I) 727-729, 1968
[10] R. Skomski, J.M.D. Coey,Permanent Magnetism, Taylor & Francis, 1999, p. 156
[11] S. Chikazumi, Physics of Ferromagnetism, 2nd ed., Oxford Science Pub., 1997, p. 252

Table I : Static permeability and resonance frequency of $(Ni_{0.40}Zn_{0.40}Cu_{0.20})_{1-c}Co_cFe_{1.98}O_4$ ferrites

| mol Cobalt | Static μ' | $f_r$ (MHz) | $μ_s×f_r$ (GHz) |
|---|---|---|---|
| Co = 0 | 270 | 22.1 | 6 |
| Co = 0.007 | 185 | 36.3 | 6.7 |
| Co = 0.014 | 160 | 45.6 | 7.3 |
| Co = 0.021 | 145 | 50.7 | 7.35 |
| Co = 0.028 | 130 | 56.5 | 7.35 |
| Co = 0.035 | 115 | 63 | |

Figure 1 : Complex permeability versus frequency of $(Ni_{0.4}Zn_{0.4}Cu_{0.2})_{1-x}Co_xFe_{1.98}O_4$ ferrites

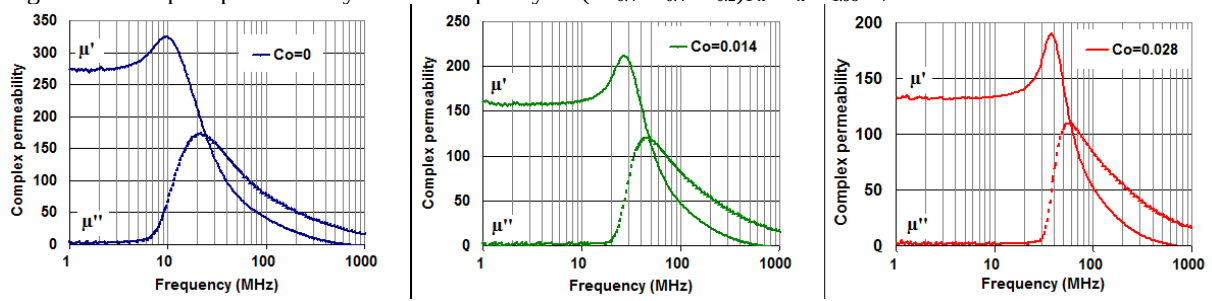

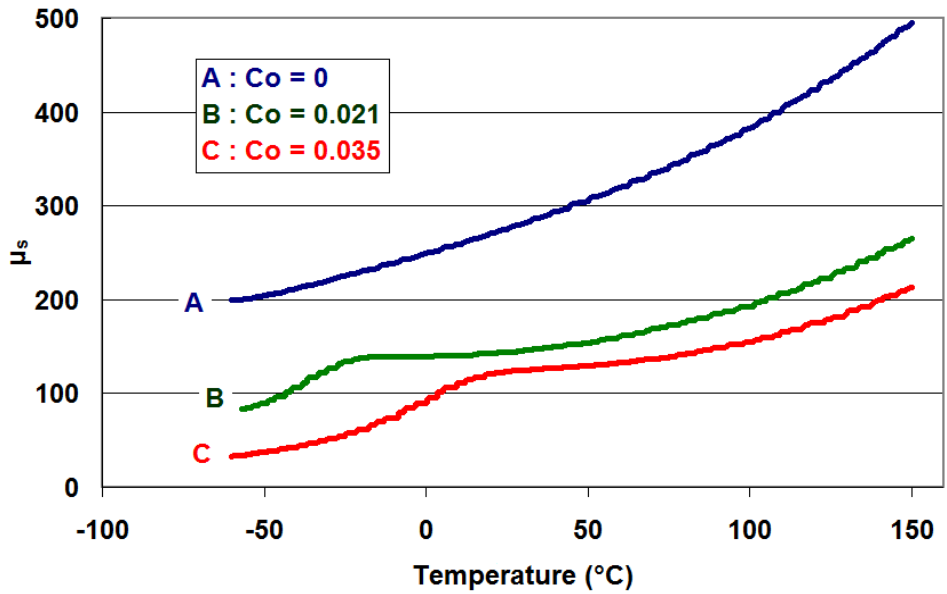

Figure 2 : Initial permeability versus temperature of $(Ni_{0.4}Zn_{0.4}Cu_{0.2})_{1-x}Co_xFe_{1.98}O_4$ ferrites.

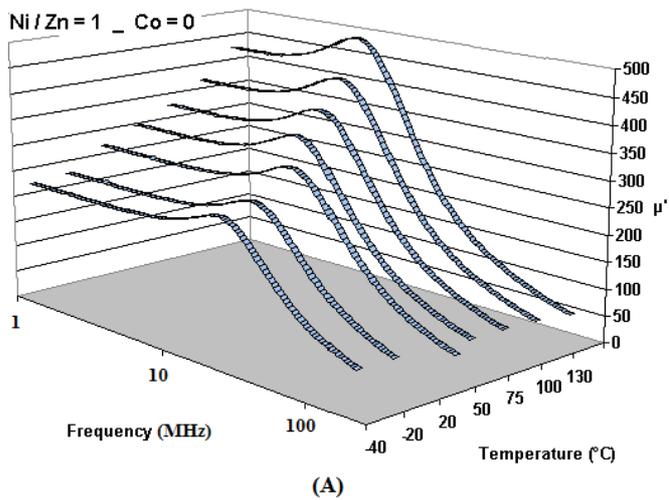

(A)

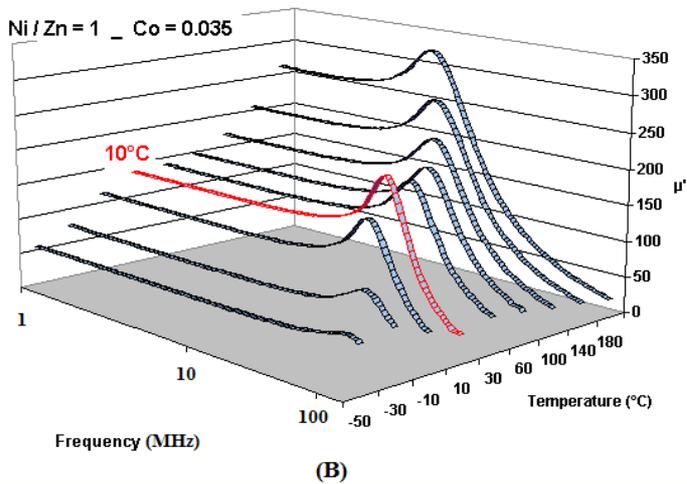

(B)

Figure 3 : (A) : $Ni_{0.4}Zn_{0.4}Cu_{0.2}Fe_{1.98}O_4$  $\mu$'(f) spectrum from 1 MHz to 110 MHz for temperature between –40°C to 130°C.

(B) : $(Ni_{0.4}Zn_{0.4}Cu_{0.2})_{0.965}Co_{0.035}Fe_{1.98}O_4$   µ'(f) spectrum from 1 MHz to 110 MHz for temperature between −50°C to 180°C.

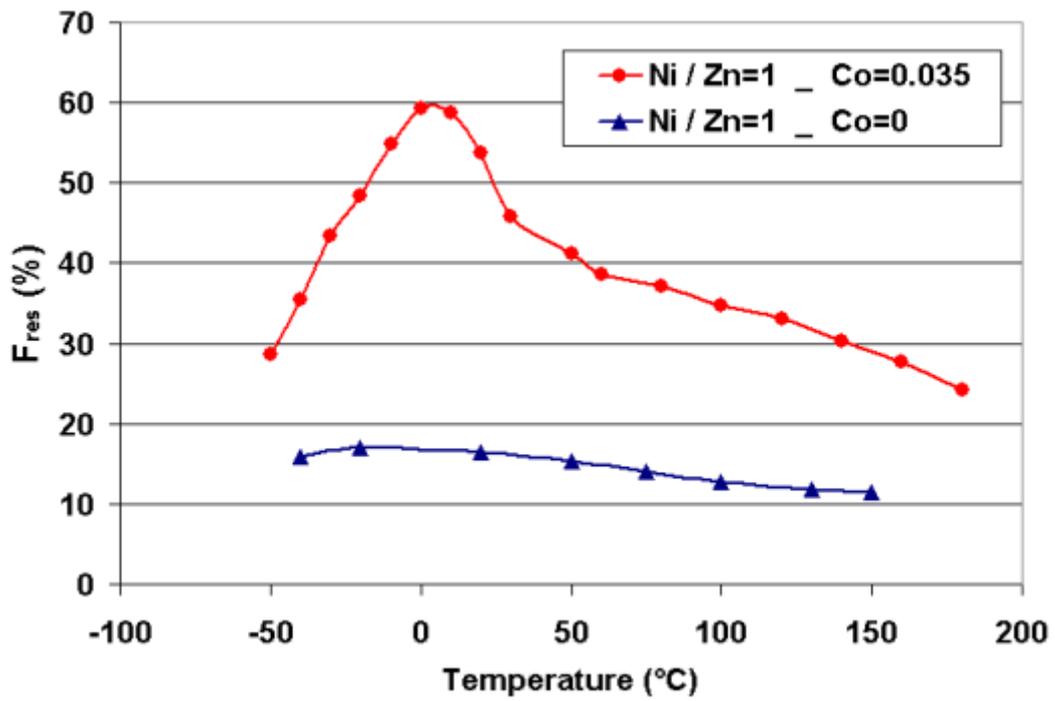

Figure 4 : $F_{res}$ versus temperature of $Ni_{0.4}Zn_{0.4}Cu_{0.2}Fe_{1.98}O_4$ and $(Ni_{0.4}Zn_{0.4}Cu_{0.2})_{0.965}Co_{0.035}Fe_{1.98}O_4$ ferrites.